# Hybrid Photonic-Plasmonic Non-blocking Broadband 5×5 Router for Optical Networks

**Shuai Sun,[1] Vikram K. Narayana,[1] Ibrahim Sarpkaya,[1] Joseph Crandall,[1] Richard A. Soref,[2] Hamed Dalir,[3] Tarek El-Ghazawi,[1] and Volker J. Sorger[1,*]**

[1]Department of Electrical and Computer Engineering, George Washington University
800 22nd Science & Engineering Hall, Washington, DC 20052, USA
[2]Engineering Department, University of Massachusetts at Boston
100 Morrissey Boulevard, Boston, MA 02125, USA
[3]Omega Optics, Inc.
8500 Shoal Creek Blvd., Bldg. 4, Suite 200, Austin, TX 78757, USA
*sorger@gwu.edu*

**Abstract:** Photonic data routing in optical networks is expected overcome the limitations of electronic routers with respect to data rate, latency, and energy consumption. However photonics-based routers suffer from dynamic power consumption, and non-simultaneous usage of multiple wavelength channels when microrings are deployed and are sizable in footprint. Here we show a design for the first hybrid photonic-plasmonic, non-blocking, broadband 5×5 router based on 3-waveguide silicon photonic-plasmonic 2×2 switches. The compactness of the router (footprint <200 µm$^2$) results in a short optical propagation delay (0.4ps) enabling high data capacity up to 2 Tbps. The router has an average energy consumption ranging from 0.1~1.0 fJ/bit depending on either DWDM or CDWM operation, enabled by the low electrical capacitance of the switch. The total average routing insertion loss of 2.5 dB is supported via an optical mode hybridization deployed inside the 2×2 switches, which minimizes the coupling losses between the photonic and plasmonic sections of the router. The router's spectral bandwidth resides in the S, C and L bands and exceeds 100 nm supporting WDM applications since no resonance feature are required. Taken together this novel optical router combines multiple design features, all required in next-generation high data-throughput optical networks and computing systems.

## 1. Introduction

The demand for higher data communication capabilities continues to rise, spanning from long haul-down to board, and even the chip level [1]. Accelerating factors beyond developments in software applications are demands for higher data capabilities in hardware implementation. However, physical limitations such as power and thermal budget constraints appose these demands restricted by technology densification as seen in multicore technology and simple I/O capacity [2]. The latter imposes restrictions on the electronic chips, known as 'dark silicon' [3]. With the bosonic nature of photons lacking a photon-photon force, data parallelism is fundamental in optics and is routinely utilized in optical data communication such as wavelength division multiplexing (WDM) [4]. With the success of long-haul optical networks, optical interconnects at the board, and even at the chip-level, have become of interest in order to mitigate the processing-to-communication gap [5]. However, the majority of optical network-on-chip (NoC) routers perform their role not exclusively in the photonic domain but often in capacitive-limiting electronics. The later also requires an overhead-heavy optic-electric-optic (O-E-O) conversion. On the other hand, one can perform routing entirely in the electronics. Yet, the known performance bottlenecks of electronic devices, namely mainly delay and power dissipation, and clamping performance. Turning to optical routing, on the other hand, is in itself inefficient given the current photonics technology due to the low light-matter interaction (LMI), and weak electro-optic modulation in silicon [6]. While photonic routers based on microring resonators have been proposed [7] and demonstrated [8], the high sensitivity (i.e. spectral and amplitude) require dynamic tunability which is both power hungry and relatively slow if high Q-factor rings are used. Hence taken together, optical routing is a) technologically cumbersome, b) latency- and energy-prone mainly due to O-E-O conversion, and c) suffers from high energy overhead due to signal error correction at the detectors TIA and laser stages, and from thermal tuning in rings-based routers [9-13].

In contrast, in this work we show an optical router design using a hybrid plasmonic-photon approach and emerging unity-high index tuning materials simultaneously to improve photonic integrated routing performance in all three factors. The enabling technological insights are based on the strong index tunability of the underlying optical plasmonic hybrid mode enabling short 2×2 switches based on voltage-controlled directional-couplers. Cascading a network of these plasmonic 2×2 switches we can design a compact optical router since the switching length scales inversely with index-change per voltage. In addition, given that the 2×2 switches are non-resonant devices due to the lossy plasmonic mode, this optical router allows for spectrally broadband operation for WDM applicability. Furthermore, unlike microrings, thermal tuning is not required, thus saving energy consumption. This hybrid photonic-plasmonic router can be synergistically deployed in Silicon-based network topology improving system performance. In our work, we use the terminology 'all-optical router' to describe the lack-of O-E-O conversion inside the router, but note that signal routing still requires electrical decision-making from the control circuit.

The remainder of this paper is organized as follows: 1) design optimization of the photonic-plasmonic hybrid 2×2 switch using indium tin oxide (ITO) as the active index modulation material. These switches are the building blocks of the router. 2) 5×5 optical router design and related operating strategies. 3) Router performance and benchmarking against existing designs.

## 2. Hybrid Photonic-Plasmonic Switch Design
### 2.1 Technology Hybridization

The fundamental building block of the optical router is a 2×2 optical switch, namely a voltage-controlled directional coupler whose performance directly impacts the overall performance of the router. Recently, photonic 2×2 switches with microring resonators (MRRs) or Mach-Zehnder Interferometers (MZIs) have been applied to perform this routing function since their spectral resonance is controlled by a voltage that changes the modal index of the ring, for instance, using the plasma dispersion effect in silicon [14, 15]. As such, the photonic MRR-based switch provides high spectral sensitivity (< 5 nm free spectral range) and low insertion loss (< 1 dB per ring). However, in order to increase the quality (Q) factor, which reduces the required ring-tuning (dynamic) power, a 10µm or even larger ring radius may be needed, which limits packaging density, and demands reasonably high power consumption during thermal tuning [7, 8]. The actual required real-estate on-chip is effectively even larger than the physical device since the electrical thermal heating pads require not only physical space but introduce thermal stray fields that need to be spread. In addition, the thermal ring response time is typically on the order of microseconds to nanoseconds, thus introducing long setup time for tuning the ring resonance [8].

To overcome the aforementioned fundamental and practical drawbacks, routing switches utilizing emerging materials beyond silicon, such as ITO, has been studied and carrier-based Drude tail modulation demonstrated [16-21]. In addition, polaritonic ('matter-like') optical modes can increase the length-scale matching between the optical-dipole moments of the gate-controlled switching materials and the optical field of the waveguide mode such as found in plasmonics [22-24]; that is, the effective group index is increased of these modes allowing for a stronger light-matter-interaction [25-27]. Incorporating those changes in the directional-coupler-based 2×2 switch enables power- and footprint-efficient switches in the following ways; the lack of a long photon-lifetime (lossy cavity and low-Q), and short carrier drift distances (~5 nm) in the index-tuning accumulation-layer inside the index-modulating ITO layer enable allow for short time responses. While a physical demonstration of the actual index tuning speed-potential in ITO is still outstanding, we estimate the carrier drift time to be sub-ps given a mobility of 15 $cm^2$/Vs for 10-20 nm thin ITO films [25]. We note that this estimation does not violate physical fundamentals, as the corresponding drift velocity is about a third of ITOs Fermi-velocity. However, based on our previous ITO experimental result in ref [25], the observed index change was an averaged value for an ITO thickness of 10 nm; meaning the actual index change is higher at near the interface, and lower further away from it [19]. That is, we double the thickness of the ITO layer (20 nm) while biasing it simultaneously from both the top and the bottom with opposite-sign voltages to achieve two accumulation layers at each ITO-insulator surface, which is beneficial for reducing the physical switch length thus enhancing the coupling efficiency discussed below. The selection for ITO as the switching material is based on its unity-strong index tunability and possible CMOS compatibility [28]. We are aware of the actual index inside the ITO film being non-homogeneous, which we explored before in ref [19]. Here, we use the experimentally-proven averaged data from ref [25] which were based on 10 nm thick ITO films. Taken together, the anticipated advantages of our 2x2 plasmonic-photonic switch are therefore a) compact physical scale, b) fast response times and short carrier drift distances to form an accumulation mode in the capacitively-gated ITO-film, and c) being spectrally broadband. While we provide a detailed loss-analysis further below for the entire 2×2 switch-based optical router, we here note that the intrinsic Ohmic plasmonic losses are actually not a detrimental factor; this is because the router is comprised of a combination of silicon photonics, namely SOI (low-loss), and plasmonics segments (high-loss), whereas the plasmonic parts are just a few micrometers in length each, thus forming a hybrid-plasmon-photon integration scheme shown in ref [29]. In fact, we find that the effective loss through this hybrid router is comparable to that through a similar-length network of silicon-based MRRs, whilst this hybrid router shows improvements in voltage, delay, and footprint and has great potentials in optical networks and even electrical-optical hybrid networks [6, 30-32].

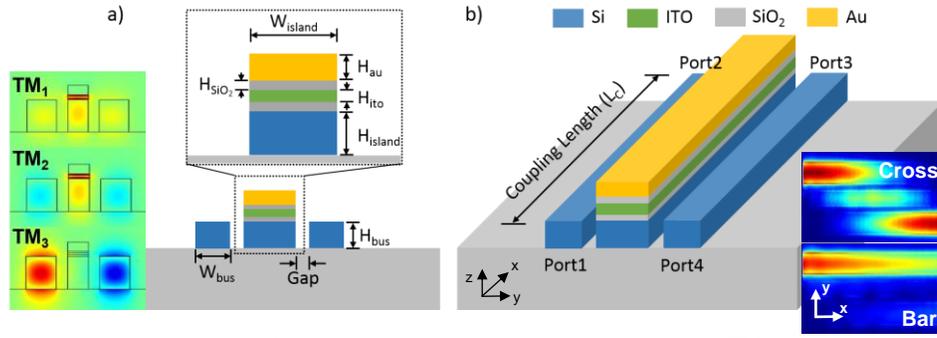

Fig. 1. Schematic design of the 2×2 hybrid photonic-plasmonic switch using ITO as the active material. The coupling length of the switch is equal to the CROSS state coupling length $L_C$. The insets are a) the $TM_1$, $TM_2$ and $TM_3$ supermodes of the 2×2 ITO switch and b) the electric filed results of the device at BAR and CROSS states at 1550 nm wavelength. The length of the ITO switch (8.9 μm) in the x-direction is not to scale. $\lambda$ = 1550 nm.

## 2.2 Switch Operating Principle

Hybridizing plasmonics with photonics reduces the propagation loss while keeping the advantages of the polaritonic optical mode [33]. Utilizing hybrid plasmon polaritons (HPPs), we added a tunable ITO layer within the metal-oxide-semiconductor (MOS) structure in order to form an electrical capacitor towards changing the optical mode's index via voltage control (Fig. 1). The switch structure includes two bus waveguides, one on each side as the input (port 1 and port 4) and the output (port 2 and port 3) ports of the switch. The center island is the actively index-tunable location of the switch. The active material is "sandwiched" between two oxide layers structure to achieve dual bias operation. The fundamental operation principle of this device is to use the index-tunable active layer (ITO layer) to switch between the CROSS state (light travels from one side of the first bus to the second bus on the other side when bias voltage $V_{bias}$ is $V_o = 0V$) and the BAR state (light stays within the bus on the same side when bias voltage is $V_{dd}$) by changing the carrier concentration of the ITO layer, thus further affecting the effective index of the supermodes governing this device; three lowest-order TM modes are spread across the cross-section of this 3-waveguide structure and can be regarded as the supermodes $TM_1$, $TM_2$, and $TM_3$ of the device (Fig. 1a). Regarding signal switching quality, we define the extinction ratio (*ER*) as the power output ratio for the BAR and CROSS port separately as its desired state (when the light is expected to be transmitted out from this port) divided by its undesired state (when the light is expected to go to the other port), Eqn (1) and (2), where port 1 is the injection port while port 2 and 3 are the BAR and the CROSS ports (in Fig.1). The insertion loss (*IL*) of the BAR and CROSS ports are defined as the power ratio between the desired port and the injection port (Eqn. 3, 4).

$$ER_{Bar} = 10\log[\frac{P_{bar-V_{dd}}}{P_{bar-V_0}}] \tag{1}$$

$$ER_{Cross} = 10\log[\frac{P_{cross-V_0}}{P_{cross-V_{dd}}}] \tag{2}$$

$$IL_{Bar} = 10\log[\frac{P_{bar-V_{dd}}}{P_{injection}}] \tag{3}$$

$$IL_{Cross} = 10\log[\frac{P_{cross-V_0}}{P_{injection}}] \tag{4}$$

$$L_B, L_C = \frac{\lambda}{2(n_{TM1} - n_{TM2})} \tag{5}$$

The coupling length difference, which is a function of the applied control bias, between the two voltage states (CROSS and BAR), needs to be maximized in order to optimize *ER* and the power consumption as well as *IL*. This leaves two design choices for an optical signal patch at the zero-voltage case: either the device is in the CROSS or BAR output state. However, since the BAR state has a longer coupling length ($L_B$) than the CROSS state ($L_C$), the physical device length of this 3-waveguide coupler is set to be the coupling length at the CROSS state. The coupling length formula for both cases is given by the difference between two symmetric TM mode indices and is related to the wavelength of the light source (Eqn. 5). While ref. [21] has shown 1.3 dB and 2.4 dB insertion losses for the CROSS and BAR switching states, respectively, two fixed values for the voltage-altered ITO effective indices. However, the Drude model for ITO allows us to select any arbitrary bias point, just limited by electrostatics such as oxide quality and contact resistance [19]. Therefore, to obtain an optimized device design, we apply the Drude model to predict the effective indices of ITO at different wavelengths [34]. Furthermore, the physical dimensions of the switch need to be optimized in order to obtain the lowest loss with the highest extinction ratios.

The insertion loss, footprint, and energy consumption of the ITO switch model in ref. [21] are already reasonably low, however, to use it as the basic element in an optical router, any small improvement of the switch *IL* are amplified by the cascaded optical router design. For example, a 0.1 dB loss reduction of a single 2×2 switch results in an over 300% (~5dB) energy savings for an 8×8 mesh network with 64 routers in total and with 8 switches in each router (calculated for the longest routing path).

## 2.3 Switch Optimization

Although the goal of the optimization is to reduce *IL* for both CROSS and BAR states while maintaining good *ER*, they cannot be improved simultaneously due to different underlying operation principles and due to the relative scaling of each of the variables; at the CROSS state, the light needs to first couple to the switching island and then to the second bus. Thus, the theoretical power that is able to be transmitted from the injection port to the CROSS port (power transmission efficiency) is critical, which determines the insertion loss ($IL_C$) at the CROSS state, and can be improved by optimizing the ratio of the island width ($W_{island}$) to the gap between the buses and the island (*Gap*) (Fig. 2a). The theoretical maximum power transmission rate (i.e. critical coupling) of the 3-waveguide coupler model at the CROSS-state occurs when the mode indices meet the condition in Eqn. 6 [35].

$$2n_{TM2} - (n_{TM1} + n_{TM3}) = 0 \qquad (6)$$

The entire optimization process follows three big steps: 1) power transmission efficiency and average loss optimization by sweeping the *Gap* and the $W_{island}$; 2) insertion loss and extinction ratio optimization by sweeping the height of the switching island $H_{island}$; 3) further performance improvement by sweeping the carrier concentration of the ITO layer.

As a first optimization step, the diameter of the bus waveguides is set to be 400 nm × 340 nm (width × height) to keep a high spatial mode confinement within the 1.4-1.7 µm single-mode operation spectrum. And $H_{island}$ is chosen to be 340 nm as an initial empirical starting point for the first optimization step. However, changing the width of the switching island also changes the $TM_1$, $TM_2$, $TM_3$ indices. Thus, changing the two variables (*Gap* and $W_{island}$) in this step also requires the thickness of both $SiO_2$ layers ($H_{SiO2}$) to be adjusted correspondingly in order to adhere to Eqn. 6. We note that there is no valid oxide thickness to satisfy Eqn. 6 beyond certain $W_{island}$ based on Lumerical MODE simulation results. Therefore, $H_{SiO2}$ is fixed to 50 nm for island widths larger than 425 nm. Here *Gap* and $W_{island}$ are swept in the range of 50~400 nm and 250~500 nm respectively, and the results are evaluated by the average channel loss of a 5×5 router, which can be regarded as a weighted metric that includes the insertion loss for both states (Eqn. 7), and the weightings depending on the router structure will be discussed in more details in the following sections.

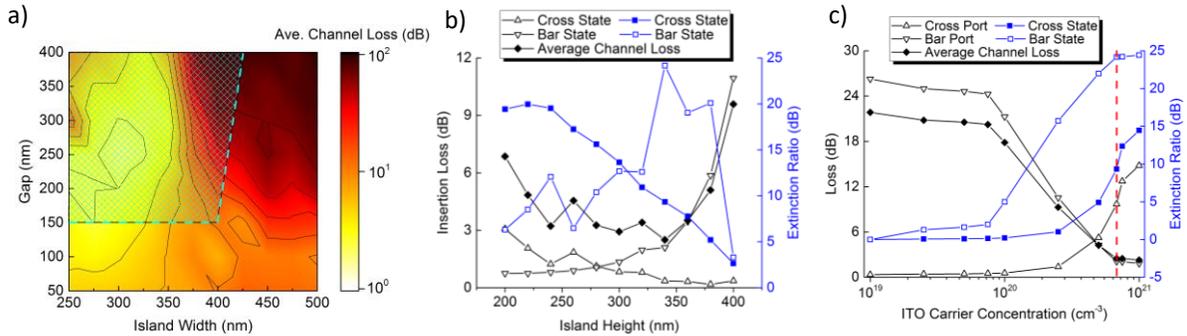

Fig. 2．2×2 switching element optimization. a) Switching island width and gap sweep for power transmission efficiency optimization; b) switching island height sweep for CROSS/BAR state insertion loss and extinction ratio trade-off; c) The ITO carrier concentrations used for the CROSS and the BAR states simulation are assumed to be $10^{19}$ cm$^{-3}$ and $6.8\times10^{20}$ cm$^{-3}$, respectively with the refractive indices 1.960 + i0.002 and 0.471 + i0.643 calculated based on the Drude model in a) and b). λ = 1550 nm.

$$Ave. Channel\ Loss = \begin{cases} 2.4 \cdot IL_{Cross} + 0.8 \cdot IL_{Bar}\ , & when\ 2IL_{Cross} \leq IL_{Bar} \\ 2.1 \cdot IL_{Cross} + 0.95 \cdot IL_{Bar}\ , & else \end{cases} \quad (7)$$

Our result shows that 2.1 dB is the lowest average channel loss at 300 nm $W_{island}$ with 250 nm *Gap* (Fig. 2a). But with denser on-chip integration, energy efficiency and high-speed design in mind, we limited the maximum device length up to 10 μm (shaded region Fig. 2a is excluded from optimization routing). By analyzing the remainder region, a 2.5 dB average channel loss is found at 275 nm $W_{island}$ with 150 nm *Gap* as the minimum loss among this sweep. Note, although the insertion loss at each state is not shown, it is important to mention that the sweet spots of insertion loss (which has an inverse relation with power transmission efficiency) at the BAR and the CROSS states distribute in different regions of the heat map. Specifically, a lower loss could be found at the bottom-left corner at the CROSS state due to shorter coupling length, while the loss at the BAR state favors larger gaps at the top because of higher $L_B/L_C$ ratio. Therefore, the average channel loss in Fig. 2a can be regarded as a loss (or power transmission efficiency) trade-off combination of two different states.

After this first step, which provides the highest power transmission efficiency in our conditions, there is still room to optimize the height of the switching island [21]. Sweeping the height of the silicon switching island $H_{island}$ below (or above) the height of the bus ('detuning') shifts the TM supermodes in Eqn. 6, and thus needs to be compensated by the thickness of oxide layers as well. Altering the $H_{island}$ from 200 nm to 400 nm, we observe a trade-off between the insertion losses of two states (Fig. 2b). This "detuning" reduces *IL* more than 10 dB at the BAR state, however, also causes an additional 2.7 dB loss at the CROSS state. After applying the average channel loss metric, the lowest loss is found at $H_{island}$ = 340 nm which happens to be the height we used in the first optimization step. We point out that this value is just for this specific 5×5 router only with weighted probabilities for two different states and a different application may result in other $H_{island}$. For example, if we assume the two states of the switch have equal probability to occur, the average loss of the switch could be reduced by 0.4 dB per switch at 240 nm $H_{island}$. In addition, the high BAR state extinction ratio is another reason to choose 340 nm $H_{island}$ without detuning.

Once the optimal island height is found, the carrier concentration of the ITO layer is the last variable that may affect the performance of the switch. We assume an experimentally proven carrier concentration change from $10^{19}$-$10^{21}$ cm$^{-3}$ as bounds for the two bias states [25]. With higher bias voltage, the carrier concentration of ITO increases due to an increased index change and eventually 'tunes' the switch to its BAR state. This can be proved by the rapid drop of *IL* at the BAR port after passing $10^{20}$ cm$^{-3}$ since there is efficient index change to make $L_B$ enough longer than $L_C$ so that the light output at the BAR state will remain in the same injection bus. It is also interesting to see that after the carrier concentration passes the epsilon-near-zero (ENZ) point (6.8×$10^{20}$ cm$^{-3}$), the average channel loss, as well as the *ER*s, saturate with little improvements. Thus, the ENZ point at 4V bias voltage is the most energy efficient BAR state, since the optical mode is here most 'spread out' given the vanishing index (i.e. strongest LMI). Our final optimized design and resulting performance parameters of the 2×2 hybrid plasmonic-photonic switch are summarized in Table 1.

Table 1 Critical design parameters and performance list of two design cases. The energy consumption is calculated based on capacitor charging energy ½ CV$^2$, and the switching time is based on device RC delay.

| Parameter | Values |
|---|---|
| Bus Diameter | 400 nm × 340 nm |
| Switch Diameter | 275 nm × 340 nm |
| Gap | 150 nm |
| ITO Height | 20 nm |
| Oxide Height | 16 nm |
| Coupling Length | 8.9 µm |
| Capacitance | 1.63 fF |
| Resistance | 500 Ω |
| Bias Voltage | 4 Volt |
| Energy per Switching | 13.1 fJ |
| Switching Time | 5.1 ps |
| BAR Insertion Loss | 2.1 dB |
| CROSS Insertion Loss | 0.4 dB |
| BAR Extinction Ratio | 24.2 |
| CROSS Extinction Ratio | 9.3 |

## 3. Hybrid Photonic-Plasmonic Router

### 3.1 Router Performance

The elemental 2×2 switches are interconnected with optical waveguides forming a switching fabric such as an *N*×*N* spatial routing switch or "matrix switch" where *N* is the number of input ports, as well as the number of output ports. For such an *N*×*N* switching network router, there are several practical architectures or layouts (Benes, Clos, etc). Here we have chosen to build the non-blocking router known as the permutation matrix, whereas the schematic design of this matrix was presented in prior works in Fig. 2(b) of ref [36], and the specific design of the matrix using 3-waveguide directional-coupler switches was given in Fig. 5 of ref [37], where this 3-waveguide design is employed here. Generally speaking, the permutation matrix has the advantage that no waveguide crossings (intersections) are used throughout in the matrix, but the matrix has the disadvantage that the overall insertion loss between an input-*i* and an output-*j* depends upon the length of the optical path traversed between the two inputs, a length that varies depending upon the specific selected *i* and *j* pair. In other words, the *IL* is path dependent.

The total number of 2×2 switches needed for a non-blocking router scales with $(N-1)^2/2$, where *N* is an odd number of ports of that router [8]. Thus, as a router for an optical mesh network of a NoC requires 4 ports to connect to the north, south, east and west neighbors, and 1 additional port for connection to the local processing core. This results in, eight 2×2 hybrid switches needed to achieve 5x5 non-blocking routing functionality that assumes assigning a random input port to a random output port without disturbing other data streams (Fig. 3). We note that other input ports are still able to maintain connections with the remainders of the output ports without affecting the initially set switches. Moreover, self-communication (communication between same input and output port number, resulting in a U-turn) is forbidden because i) it can be achieved with higher energy- and latency- efficiency with other local (electrical) interconnect links, and ii) avoiding self-communication can simplify the router from $N^2$ number of switches required for all-to-all connection down to only $(N-1)^2/2$, which can also reduce the average loss of the router.

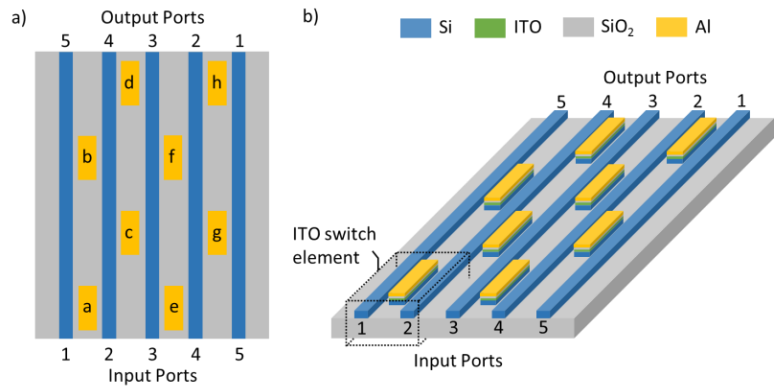

Fig. 3. The top view and the schematic plot of the 5×5 Port non-blocking optical router. 8 individual 2×2 ITO switches are placed with certain pattern in order to achieve non-blocking routing function. The length of the ITO switches are not to scale.

The operational spectrum results for each output port with respect to cross-coupling from other routing paths are key parameters for signal quality and to assess the WDM ability (Fig. 4). For example, configuring the router to establish the following paths: 1 to 2, 2 to 3, 3 to 4, 4 to 5 and 5 to 1, and injecting a unity laser power ($P_{laser}$ = 100% a.u.) from port 1, results in the majority of the signal to be routed to port 2, as designed while the leakage is delivered to the remaining four output ports. The 3 dB spectral (not temporal) bandwidth, i.e. routed signal dropping to -3dB from maximum, is 106 nm wide on average for all 20 different routing paths (130 nm from 1.49 to 1.62 µm, Fig. 4a). The broad bandwidth with an average signal-to-noise ratio (SNR) of 123 resulting in an average channel capacity of 10×5 Gbps (10×6 Gbps in Fig. 4a due to above average bandwidth) per routing path based on CWDM standard across the S, C and L bands with 20 nm wavelength spacing (Fig. 4a) and 200 Gbps in total if all five ports are used. Here, the SNR is defined as the power ratio between the signal and the light leakage to the other ports. Furthermore, the data capacity can be improved by using DWDM in C band (1530~1560 nm wavelength) with 0.8 nm wavelength spacing which supports 40 wavelengths and results in 400 Gbps data capacity per channel (Fig. 4b). Note, this data capacity is calculated based on the standard of the 10 Gigabit Ethernet with 10 Gbps data rate per wavelength [38]. However, the ideal Shannon data capacity based on the device 3 dB bandwidth and average SNR is about 92 Tbps based on Eqn. 8, which shows the maximum capacity of a single routing path with advanced coding strategies such as PAM, QAM, and PWM, etc [39]. We note that this router is WDM capability in that is it supports multiple wavelengths per light path. While individual wavelength routing is not possible, multiple pre-multiplexed wavelength channels could be routed jointly, and post-routing demultiplexed.

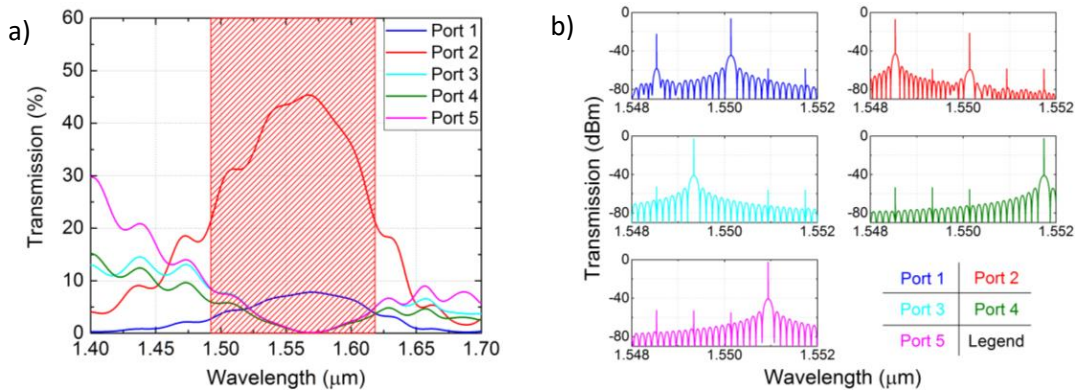

Fig. 4. Router performance simulation. The router is configured to route the signal from each port to the next one (i.e. port 1 to port 2, port 2 to port 3, etc.). a) Single-wavelength-single-input from port 1 for operation spectrum testing; b) five-wavelength-five-input with each input port assigned to a wavelength for WDM testing with 0.8 nm wavelength spacing. The shaded area in a) represents the 3dB bandwidth which covers from 1.49 µm to 1.62 µm wavelength range.

Doing so increases the data capacity of this particular circuit-switched path by a factor equal to the number of wavelengths used (e.g. 100). This could be exploited in applications such as optical residue computing or optical reduction operations.

$$Shannon\ Capacity = BW \cdot log_2(1 + SNR) \qquad (8)$$

The port-to-port crosstalk is tested by injecting five light source in five different wavelengths and we find that the port-to-port crosstalk is at least -13 dB higher than the signal power received by other ports (Fig. 4b). Interestingly, different from ring-based WDM optical routers that only support one wavelength at a given time window, the WDM ability of this router allows for multiple wavelengths to be supported simultaneously with no thermal resonance tuning needed. Moreover, the average performance for all 20 different paths is concluded in the next section.

### 3.2 Operation Strategy

A single 2×2 ITO switch does not consume any active (voltage-driven) energy in its CROSS state operation since the bias signal equals to zero. Here, the ITO layer has dielectric-like properties and exhibits low insertion loss in this 3-waveguide coupler structure just like passive silicon couplers. However, applying the bias voltage changes the ITO layer to its metallic state and the coupling length increases accordingly. Although light cannot be coupled to the bus on the other side due to insufficient coupling length, the high-loss plasmonic mode inflicts some amount of BAR state loss since the optical mode 'bounces' back from the metallic center island (non-zero interaction with the plasmonic mode). As a result, the BAR-states' insertion loss, $IL_B$, is higher than that for the CROSS state. The loss difference of the BAR and CROSS states of the router allows reconfiguration to reduce the overall routing loss by choosing a path routing with an increases number of CROSS/BAR switching events, when able. For example, to route a signal from input port 3 to output port 4, all the possible routing paths with switch states are listed in Table 2 and the one with more CROSS state and less BAR state provides the lowest routing loss. Following this routing strategy, the average channel loss for all 20 different routing paths can be reduced to 2.5 dB for single routing path with 1.1 dB as the best case, and 3.2 dB as the worst case. Moreover, a comparison between this hybrid router and other photonic routers is summarized in Table 3. Note, even though the response time of the router mainly depends on RC delay of the switch itself, we limited the switching speed up to 10 GHz which is commonly accepted in the optical communication community with the concern of heat dissipation and energy efficiency [1, 40].

Table 2 Routing path options from port 3 to port 4 of the theoretical ideal case. The router states from left to right represents the switch states from switch *a* to switch *h*.

| Path Options | \\ Switch States \\ | | | | | | | | Loss (dB) |
|---|---|---|---|---|---|---|---|---|---|
| | *a* | *b* | *c* | *d* | *e* | *f* | *g* | *h* | |
| 1 | - | Bar | Cross | Bar | Bar | - | - | - | 6.6 |
| 2 | - | - | - | Cross | Cross | Cross | Bar | - | 3.1 |
| 3 | - | - | Bar | Cross | Bar | Bar | - | - | 6.6 |

Table 3 Parameter comparison among this hybrid photonic-plasmonic router and other photonic router designs. MZI, MRR, and IPS stand for Mach-Zehnder interferometer, micro-ring resonator, and hybrid photonic-plasmonic switch. The projects of Li [15], Ji [7], Yaghoubi [41] and Jia [8] are results from fabricated and tested devices, while Dang [42] and this work are results based on numerical simulations.

| Project | X. Li (2013) | R. Ji (2013) | D. Dang (2015) | E. Y. (2016) | H. Jia (2016) | This Work |
|---|---|---|---|---|---|---|
| Key Element | MZI | MRR | MRR | MZI | MRR | HPPS |
| Element Number | 10 | 16 | 16 | 20 | 8 | **8** |
| Single Wavelength Data Capacity (Gbps) | 32 | 13 | 40 | 20 | 32 (1280 for WDM) | **50 (250~2000 for WDM)** |
| Energy Consumption (fJ/bit) | 781* | - | 0.4 (fW/bit) | 1442* | 68.2 | **5.2 (1.0~0.1 for WDM)** |
| Average Loss (dB) | 2.4 | **1*** | 1.6* | 6.0* | 16.5 | 2.5 |
| Maximum Loss (dB) | 9.6 | - | **2.4*** | 8.4* | 18.3 | 3.2 |
| Area (um$^2$) | 9.6×10$^5$ | 4.6×10$^5$ | - | - | 4.8×10$^5$ | **200** |
| 3dB Bandwidth (nm) | 40 | 0.4 | ** | 100 | 0.6 | **106** |
| SNR | 24 | 10 | - | 34 | 11 | **123** |
| Switching Time (ps) | 10$^6$/10$^3$*** | - | **100** | - | 2×10$^7$ | **100** |

\* Numbers are not directly given and calculations or approximations are used to obtain the values.
\*\* This device only operates at two certain wavelengths: 1547.5 and 1550 nm.
\*\*\* This device allows both slow (μs) and fast (ns) switching by thermal and electrical tuning.

## 4. Conclusion

In summary, we designed and analyzed a hybrid photonic-plasmonic non-blocking broadband on-chip router on a Silicon photonics platform. The router response time (0.1 ns) and high-energy efficiency (1.0 and 0.1 fJ per switching for CWDM and DWDM respectively) are enabled by hybridizing plasmonics with a photonic device. In comparison with microring- and Mach Zehnder-based photonic routers, this router operates over a broadband 3-dB signal discrimination bandwidth over 100 nm allowing up to 2 Tbps theoretical noisy Shannon channel capacity. The design is enabled by a hybrid photonics-plasmonic integration strategy featuring cascaded 3-waveguide-based 2x2 switches, utilizing ITO's strong voltage-controlled index tunability. Using these plasmonic switches allows compact router designs of 200 μm$^2$ footprint and 10$^2$ times area-utilization improvement. The high performance and scalability of this router are promising features for large-scale multi-core optical networks requiring all-optical routing applications.

## Acknowledgements

V.S. and T. E. are supported by the Air Force Office of Scientific Research (AFOSR) award number FA9550-15-1-0447 which is part of the Dynamic Data-Driven Applications System (DDDAS) program, and V.S. by AFOSR award number FA9550-14-1-0378. H.D. and V.S. are supported by AFOSR (FA9550-17-P-0014) of the small business innovation research (SBIR) program.